\documentclass[twocolumn,showpacs,preprintnumbers,amsmath,amssymb,floatfix]{revtex4}

\usepackage{graphicx}
\usepackage{epsfig}
\usepackage{dcolumn}
\usepackage{bm}

\begin{document}


\title{A new measurement of the altitude dependence of the atmospheric muon intensity}

\author{J.J. Beatty,\footnote{Present address: Department of Physics, Ohio State 
University, Columbus, OH 43210} S. Coutu, and S.A. Minnick,\footnote{Present 
address: Department of Physics, Kent State University Tuscarawas Campus, 
New Philadelphia, OH 44663}\footnote{Corresponding author: sminnick@kent.edu}}
\affiliation{Departments of Physics and of Astronomy and Astrophysics, 104
Davey Laboratory, \\
The Pennsylvania State University, University Park, PA 16802}%

\author{A. Bhattacharyya, C.R. Bower, and J.A. Musser}
\affiliation{Department of Physics, Swain Hall West, Indiana University,
Bloomington, IN 47405}%

\author{S.P. McKee, M. Schubnell, G. Tarl\'{e}, and A.D. Tomasch}
\affiliation{Department of Physics, Randall Laboratory, University of
Michigan, \\
500 E. University Ave., Ann Arbor, MI 48109-1120}%

\author{A.W. Labrador,\footnote{Present address: Department of Physics,
California
Institute of Technology, Pasadena, CA 91125}
D. M\"{u}ller, and S.P. Swordy}
\affiliation{Enrico Fermi Institute and Department of Physics, 933 E.
56$^{th}$ St., \\
University of Chicago, Chicago, IL 60637}%

\author{M.A. DuVernois}
\affiliation{School of Physics and Astronomy, 16 Church St. SE, \\
University of Minnesota, Minneapolis, MN 55455}%

\author{S.L. Nutter}
\affiliation{Department of Physics and Geology, SC 147, \\
Northern Kentucky University, Highland Heights, KY 41099}%

\date{\today}

\begin{abstract}
We present a new measurement of atmospheric muons made during an ascent 
of the High Energy Antimatter Telescope balloon experiment.  The muon 
charge ratio $\mu^{+}/\mu^{-}$ as a function of atmospheric depth in the
momentum interval 0.3--0.9 GeV/c is presented.  The differential $\mu^{-}$
intensities in the 0.3--50 GeV/c range and for atmospheric depths between 
4--960 g/cm$^{2}$ are also presented.  We compare these results with other
measurements and model predictions.  We find that our charge ratio is 
$\sim$1.1 for all atmospheric depths and is consistent, within errors, 
with other measurements and the model predictions.  We find that our 
measured $\mu^{-}$ intensities are also consistent with other measurements, 
and with the model predictions, except at shallow atmospheric depths.

\end{abstract}


\maketitle

\section{Introduction}

Measurements and theoretical calculations of atmospheric neutrino fluxes
have consistently disagreed, and this disagreement was interpreted in 
terms of neutrino oscillations. However, this interpretation requires 
an accurate understanding of neutrino fluxes in the atmosphere.  
Simulations of the absolute fluxes have been conducted 
(\cite{barr:muons, wentz:muons, gaisser:muons}, and 
references therein) but suffer from systematic uncertainty in the 
normalization of the atmospheric neutrino spectrum.  These simulations
also predict the spectrum of other atmospheric secondaries, 
specifically muons.  Indeed, the production of neutrinos in the
atmosphere is closely coupled with that of muons, as they are produced
together in pion and kaon decays, and as some muons themselves decay
and contribute to the neutrino flux. Therefore, measurements of 
atmospheric muons by ascending high altitude balloon borne instruments
can be used to reduce the neutrino model uncertainties. It should be
pointed out that other experimentally accessible quantities can also 
be used to that end, such as the absolute primary cosmic ray flux 
impinging on the atmosphere, or other atmospheric secondaries such 
as gamma rays or secondary nucleons. The atmospheric muon flux is 
therefore an important element in any comprehensive strategy to
improve our understanding of neutrino production in the atmosphere.
  
An earlier version of 
the High Energy Antimatter Telescope (HEAT-e$^\pm$) instrument had been used to 
measure air-shower muons during atmospheric ascent~\cite{coutu:muons}.  
The HEAT instrument is described elsewhere~\cite{barwick:heat, beach:heat}.  
In its present configuration, HEAT-pbar is optimized to study antiprotons.  
It combines a superconducting magnet spectrometer 
using a drift-tube hodoscope (DTH), a time-of-flight system (TOF), 
and two stacks of multiwire proportional chambers (dE/dx). We report here a new 
measurement of the muon charge ratio $\mu^{+}/\mu^{-}$ as a function of 
atmospheric depth in the momentum interval 0.3--0.9 GeV/c, and
differential $\mu^{-}$ intensities in the 0.3--50 GeV/c range and for atmospheric 
depths between 4--960 g/cm$^{2}$ for a balloon flight from Fort Sumner, NM, USA 
on June 3, 2000.

\section{Muon Identification and Backgrounds}

The ascent phase of the flight lasted $\sim$3 hours during which more 
than 1.5 million events were recorded at an average vertical geomagnetic
rigidity cutoff of 4.5 GV.  The flight occurred under solar maximum 
conditions ($\Phi$ = 1330 MV).  The atmospheric overburden ranged from a high 
of 960 g/cm${^2}$ with the instrument on the ground to a low of 4 g/cm${^2}$ at 
float altitude.

Muon events were identified using the three main detectors onboard the 
HEAT-pbar instrument.  The methods of muon identification and selection
criteria involving the TOF and DTH detectors are described in detail 
elsewhere~\cite{coutu:muons}.  To select muon events using the TOF 
detector, the incident particle must have $0.85 \leq \beta \leq 2.00$,
where $\beta = v/c$ is the particle's velocity in units of the speed of
light.  Also, for $\mu^{+}$, the particle must have 0.3 GV $<$ R $<$ 0.9 GV and 
for $\mu^{-}$, R $<$ -0.3 GV, where ${\rm R}=pc/Ze$ is the particle's 
magnetic rigidity, or momentum per unit electric charge.
Muon events with R $>$ 0.9 GV cannot be selected based on rigidity as a 
function of $\beta$, due to proton contamination.  The 
range of $\beta$ selected also ensures that the particle was traveling 
downward through the instrument.  To select muon events with the DTH 
detector, the particle track must have passed through at least 4 of the 
8 drift tube planes in the non-bending view, and at least 8 of the 18
planes in the bending view.  This is to ensure that there are enough samples 
along the particle's track to produce a valid track and rigidity measurement 
through the DTH.  
We retain only events with average fit residuals along the track 
that are better than 6~mm in the x direction, 1.5~mm in the y direction, and 
1~mm in the z direction. This rejects only $\sim$0.1\% of events, and provides 
basic track cleanliness criteria.
Lastly, the event is required to have a tracking fit goodness parameter
$\chi^{2}_{DT} \leq$ 10.0 and $\mid$(Maximum Detectable Rigidity)/R$\mid$ $\geq$ 4, 
which are used to ensure good track quality.

\subsection{The dE/dx Detector}

\subsubsection{Physical Construction}

\begin{figure}[ht]
\centering
\epsfxsize=3.0in
\epsfbox{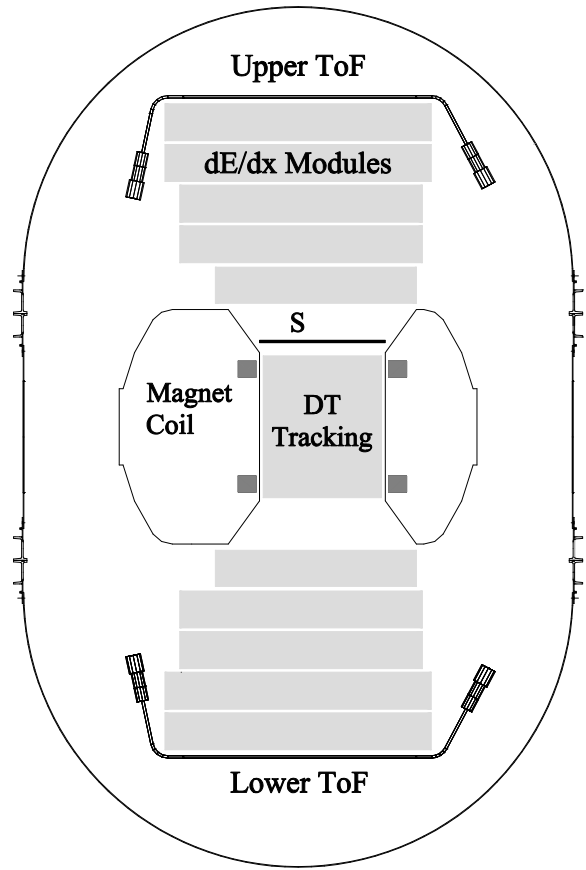}
\caption{Schematic view of the HEAT-pbar Instrument.  The solid black 
line labeled ``S'' is a piece of scintillator and is part of the 
instrument trigger.}
\label{fig:instr}
\end{figure}

The dE/dx detector consists of two stacks of 70 segmented multi-wire 
proportional chambers, one stack above and one below the magnetic 
hodoscope as shown in Figure~\ref{fig:instr}.  The 140 chambers are 
each $\sim$1 cm thick and filled with a gas mixture of 95\% Xenon and 
5\% Methane (used as a quench gas).  Xenon was used since it yields 
the strongest possible logarithmic rise in the energy loss curve from 
minimum ionization to relativistic saturation (70\%), and hence gives 
the highest possible velocity resolution.

\subsubsection{Energy Loss Determination}

\begin{figure}[ht]
\centering
\epsfxsize=3.0in
\epsfbox[80 200 505 635]{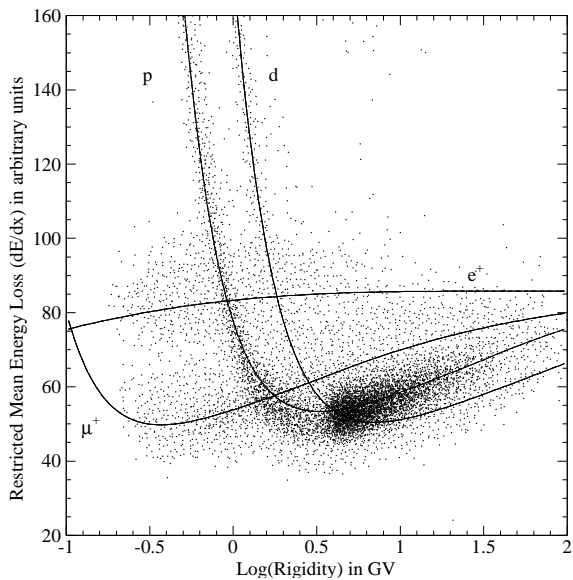}
\caption{Distribution of dE/dx as a function of rigidity for ascent 
data for positive rigidities.  Superimposed are the Bethe-Bloch energy
loss 
curves for various particle species.}
\label{fig:dedx}
\end{figure}

\begin{figure}[ht]
\centering 
\epsfxsize=3.0in
\epsfbox[80 200 515 635]{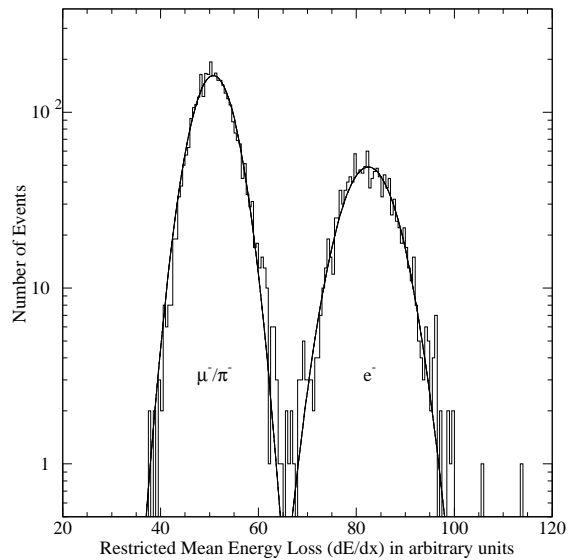}
\caption{Restricted mean energy loss distributions for ascent data with
gaussian 
fits for -0.9 GV $<$ R $<$ -0.3 GV.}
\label{fig:dedx_gauss}
\end{figure}

Each chamber's anode wires are connected in groups of 4-6 wires.  
In addition to the ionization energy deposits, there are occasional spurious
signals in wire groups away from the particle track. For analysis,
only wire groups less than 5 cm (the average
width of two adjacent wire groups) from the particle's track are used in
the energy loss determination for each chamber.  The signals are corrected
for detector response changes during the flight, electronic gain variations, 
and particle incident angle. 
Some particle trajectories are within the instrumental acceptance but miss 
some of the chambers near their edges, so that the mean number of chambers hit 
in an event is 128 out of 140.
We require the particle to have left a signal in at least 103 of 140 
chambers to ensure sufficient sampling points for a determination of the 
average energy loss. The restricted mean average energy loss is
determined by eliminating chambers whose signal is not in the lowest 58\% of
the signals from all chambers hit.  
The restricted mean average is used 
to minimize the effects of Landau fluctuations on the final energy loss 
determination.  A typical plot of restricted mean energy loss (dE/dx) 
versus positive rigidity for ascent data is shown in Figure~\ref{fig:dedx}.  
Included on the plot are the Bethe-Bloch energy loss curves for deuterons,
protons, muon/pions, and positrons. Various particle species clearly
populate the plot. Note the geomagnetic cutoff effect on the proton data at 
$\sim$4.5 GV (Fort Sumner, NM, USA).
Figure~\ref{fig:dedx_gauss} shows the restricted mean energy loss for
events in the rigidity range  -0.9 GV $<$ R $<$ -0.3 GV, as an example, which
highlights the highly gaussian nature of the restricted mean energy loss.
Distributions in other rigidity ranges look similar.

\subsubsection{dE/dx Event Selection}

\begin{figure}[ht]
\centering
\epsfxsize=3.0in
\epsfbox[80 200 515 635]{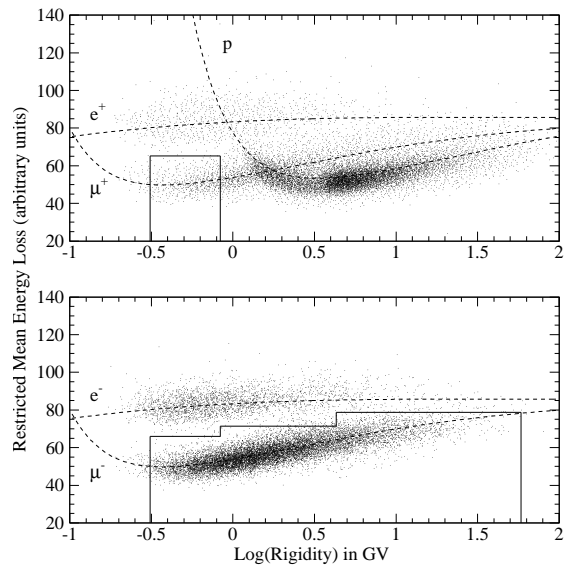}
\caption{Restricted mean energy loss as a function of rigidity for 
ascent data for \textit{(top)} positive rigidities and \textit{(bottom)}
negative rigidities.  The event selections (solid lines) are shown on each
plot as well as the Bethe-Bloch energy loss prediction curves (dashed).  
Note that the apparent higher density of negative muons compared to positive 
muons is an artifact of the plotting routine used for the figure; there are
approximately equal numbers of points shown on each panel.}
\label{fig:dedx_v_r}
\end{figure}

For the dE/dx detector, the difference between the restricted mean energy
loss for the upper and lower dE/dx stacks must be within approximately 
two standard deviations from the mean of the difference distribution.  
This ensures that an event that showers below the top dE/dx stack is
rejected.  
The restricted mean dE/dx signal of the particle must fall between 
0 and 65 (inclusive, in arbitrary units) for 0.3 GV $< |$R$| <$ 0.8 GV, 
between 0 and 70 for 0.8 GV $< |$R$| <$ 4.0 GV, and between 0 and 78 for
4.0 GV $< |$R$| <$ 50.0 GV.  See Figure~\ref{fig:dedx_v_r} for a plot of
restricted mean energy loss versus rigidity.

\subsection{Backgrounds to the Differential Muon Intensity}

Protons, electrons, and pions are the major background contaminations 
of the differential muon intensity.  The pion intensity, though relatively small 
with respect to the muon intensity, increases with altitude.  From 
\cite{stephens:pions}, the pion/muon ratio is $\sim$0.035 at a depth of 3 g/cm$^2$.  
Fortunately, the intensities of other products of atmospheric 
showers, such as kaons, are small in relation to the muon intensity.

Particle interactions within the HEAT-pbar instrument also produce pions
that can contaminate the muon intensity.  A GEANT 3.21 Monte Carlo simulation
of the instrument with an input proton spectrum falling with an index of
$-2.7$ and an input momentum range of 4.5 GV $\leq$ R $\leq$ 50 GV yielded
a secondary interaction pion to proton fraction of $1.1 \times 10^{-3}$,
which decreased to $9.7 \times 10^{-5}$ after application of the muon 
selection criteria.  In order to determine what percentage of the muon 
intensity this represents, and hence the contamination percentage, the proton
to muon ratio must be known. As a first order estimate, we used proton
and muon information from \cite{grieder:CRearth}, and determined that the 
proton-to-muon ratio ranges from $1.6 \times 10^{-4}$ at the ground to a
maximum of $3.0$ at 13 g/cm$^{2}$.  
Choosing the highest value for the ratio offers a worst case contamination
fraction, since the ratio decreases with depth.  Using a ratio of 3.0 
yielded an interaction pion intensity that was 0.01\% of the proton intensity 
and 0.03\% of the muon intensity at atmospheric depths greater than 13 g/cm$^2$. 
These percentages only decrease further with atmospheric depth, so the 
interaction pion contamination of the muon intensity is ignored since the error
introduced is much less than the statistical error of the measurement.  
The number of kaons produced in the instrument is even smaller than the 
pion number, so they are ignored as well.

\section{Results and Discussion}

\subsection{Muon Charge Ratio}

\begin{figure}[ht]
\begin{center}
\includegraphics*[2.25in,4.0in][5.75in,7.25in]{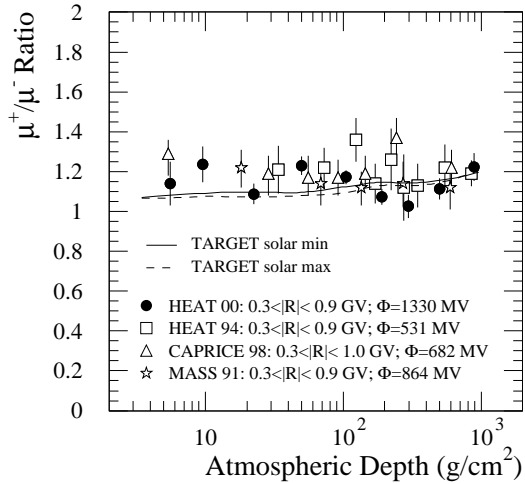}
\end{center}
\vspace{-0.51in}
\caption{Atmospheric muon charge ratio as a function of atmospheric depth for 
a geomagnetic cutoff of 4.5 GV (Fort Sumner, NM, USA).  Also shown are the 
ratios for the HEAT 1994 flight as well as other measurements.  The curves 
are calculations with the 3-dimensional TARGET algorithm. Indicated are the 
solar modulation parameters for each data set.}
\label{fig:ratiofs}
\end{figure}

\begin{figure}[ht]
\begin{center}
\includegraphics*[2.25in,4.0in][5.75in,7.25in]{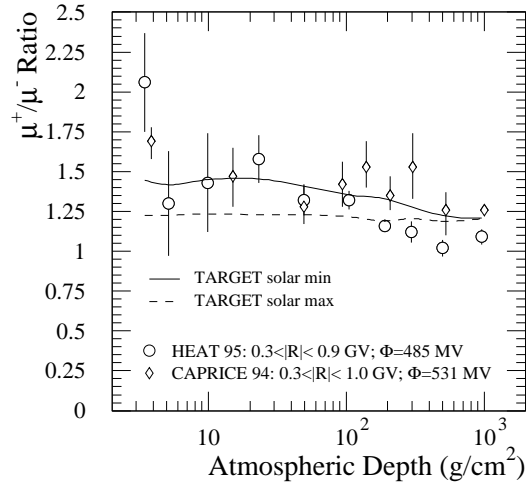}
\end{center}
\vspace{-0.51in}
\caption{Atmospheric muon charge ratio as a function of atmospheric depth for 
a geomagnetic cutoff of $\ll$ 1 GV (Lynn Lake, Manitoba, Canada) for two data 
sets.  The curves are calculations with the 3-dimensional TARGET algorithm. 
Indicated are the solar modulation parameters for each data set.}
\label{fig:ratioll}
\end{figure}

The number of muons for various atmospheric depths for the rigidity 
range 0.3 $\leq$ $\mid$R$\mid$ $\leq$ 0.9 GV is shown in Table~\ref{tbl:ratio}.
\begin{table}[ht]
\caption{Muon charge ratio as a function of atmospheric depth (d) for 
0.3 $\leq$ $\mid$R$\mid$ $\leq$ 0.9 GV.  Errors on the ratios are statistical
only.}
\begin{center}
\begin{tabular}{cccccc}
\hline\hline
d & $\overline{d}$ & $N_{\mu^{+}}$ & $N_{\mu^{-}}$ & 
$N_{\mu^{+}}/ N_{\mu^{-}}$ \rule{0mm}{4mm} \\
$(g/cm^{2})$ & $(g/cm^{2})$ &  &  & \\
\hline
4--7 & 5.572 & 228 & 200 & 1.14 $\pm$ 0.11 \rule{0mm}{4mm} \\
7--13 & 9.583 & 428 & 346 & 1.237 $\pm$ 0.089 \\
13--32 & 22.49 & 910 & 837 & 1.087 $\pm$ 0.052 \\
32--67 & 49.68 & 1539 & 1251 & 1.230 $\pm$ 0.047 \\
67--140 & 105.2 & 2460 & 2099 & 1.172 $\pm$ 0.035 \\
140--250 & 190.8 & 1552 & 1447 & 1.073 $\pm$ 0.039 \\
250--350 & 296.7 & 658 & 642 & 1.025 $\pm$ 0.057 \\
350--840 & 497.5 & 949 & 852 & 1.114 $\pm$ 0.053 \\
840--960 & 887.3 & 688 & 562 & 1.224 $\pm$ 0.070 \\
\hline\hline
\end{tabular}
\end{center}
\label{tbl:ratio}
\end{table}
The $\mu^{+}/\mu^{-}$ ratio is shown as a function of atmospheric depth in
Figure~\ref{fig:ratiofs} for a geomagnetic cutoff of $\sim$4.5 GV (Fort Sumner, 
NM, USA), and in Figure~\ref{fig:ratioll} for a very small cutoff at northern 
latitudes (Lynn Lake, Manitoba, Canada).  Shown are the new HEAT-pbar 
measurements (labeled HEAT 00), as well as two previous HEAT-e$^\pm$ measurements 
and other recent measurements (\cite{coutu:muons, boezio:muons, bellotti:muons}).  
The various measurements were made at different solar 
epochs, so that direct comparison is problematic.  However, there is good agreement 
between the various measurements at Fort Sumner, irrespective of the solar 
modulation conditions, which is to be expected since the primary flux modulation 
is small above 4.5 GV.  There are only two measurements at Lynn Lake, with 
relatively large uncertainties, and where solar modulation effects are expected 
to be more important.  Both measurements were made under near solar maximum 
conditions, and are compatible with each other given the large uncertainties.  
Also shown on Figures~\ref{fig:ratiofs} and~\ref{fig:ratioll} are 3-dimensional 
calculations of the air shower development with the TARGET 
algorithm~\cite{agrawal:target}, widely used for neutrino flux calculations.  
Curves are shown for both solar maximum and minimum conditions, for average 
primary fluxes (which may not represent the actual spectrum at the time of any 
individual measurement).  The agreement between the calculations and the 
measurements (especially the HEAT measurements) appears reasonably good, with 
the caveat that statistical uncertainties at high altitudes for the HEAT 95 
measurements are large.

\subsection{$\mu^{-}$ Differential Intensity}

\begin{figure}[ht]
\includegraphics*[2.5in,4.2in][5.8in,7.1in]{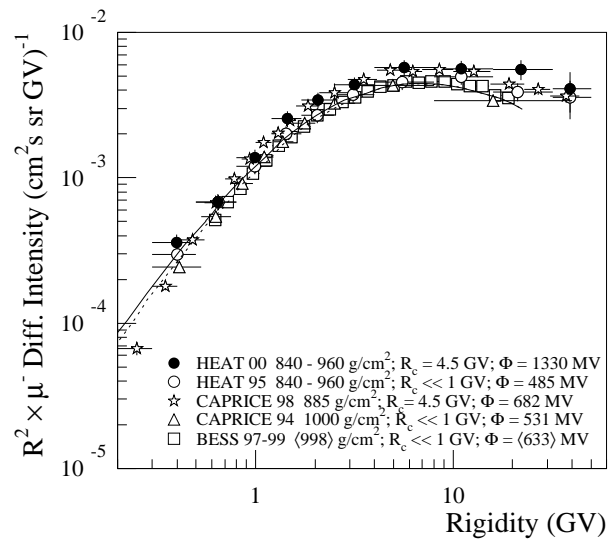}
\caption{\label{fig:flux_gnd}
Ground level differential $\mu^{-}$ intensities as a function of rigidity.  Also 
shown are results from various other instruments.  The plots have been scaled by 
a factor of $R^{2}$.  The curves are calculations with the 3-dimensional TARGET 
algorithm for solar minimum (\textit{solid}) and solar maximum (\textit{dashed}).}
\end{figure}

\begin{figure*}
\includegraphics*[2.4in,2.8in][5.8in,8.5in]{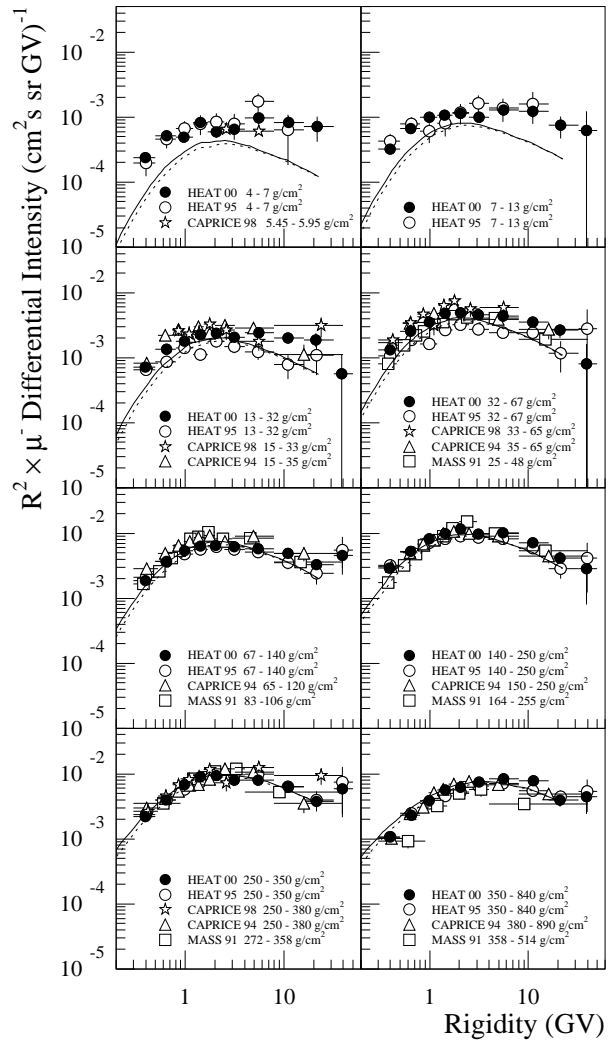}
\caption{\label{fig:flux_atm}
Differential $\mu^{-}$ intensities as a function of rigidity for 
various atmospheric depth ranges.  Also shown are results from various 
other instruments. (Note: only subsets of measurements by other groups are shown 
here, where the choice of atmospheric depth bins matches with ours.)  The plots 
have been scaled by a factor of $R^{2}$.  
The curves are calculations with the 3-dimensional TARGET algorithm for 
solar minimum (\textit{solid}) and solar maximum (\textit{dashed}).  See
Figures~\ref{fig:ratiofs} and~\ref{fig:ratioll} for each measurement's solar and 
geomagnetic conditions.}
\end{figure*}

\begin{figure*}
\includegraphics*[2.4in,2.8in][5.8in,8.5in]{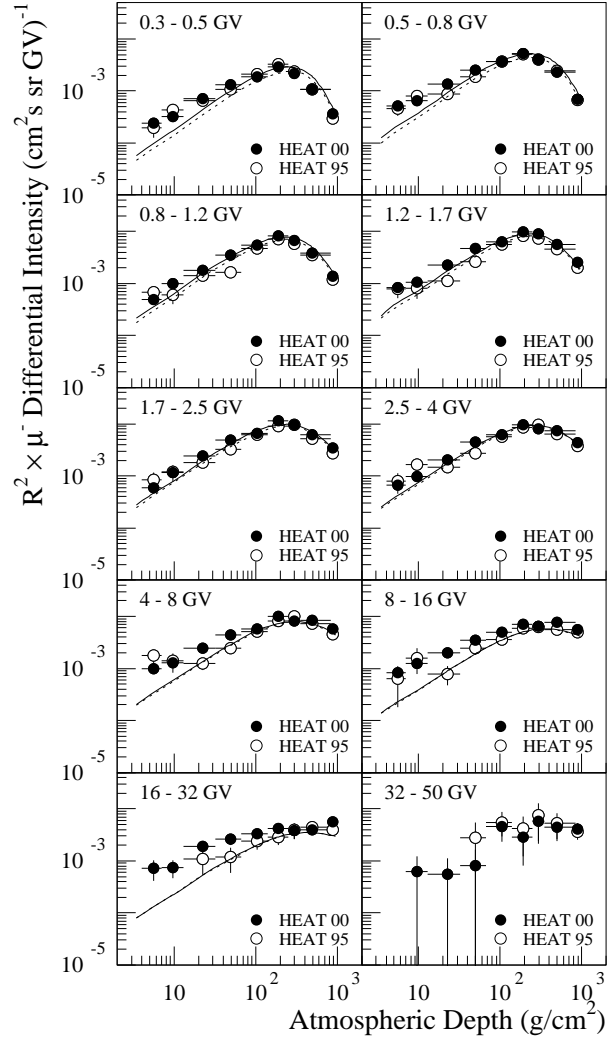}
\caption{\label{fig:growth}
$\mu^{-}$ growth curves for various rigidity intervals.  (Note: other groups 
have used different rigidity bins in their analyses, so that we only report 
HEAT measurements here.)  The plots have been scaled by a factor of $R^{2}$.  
The curves are calculations with the 3-dimensional TARGET algorithm for 
solar minimum (\textit{solid}) and solar maximum (\textit{dashed}).
See Figure~\ref{fig:flux_gnd} for each measurement's solar and 
geomagnetic conditions.}
\end{figure*}

Our measurement of the $\mu^{-}$ differential intensities was made using 
similar corrections as for previous HEAT-e$^\pm$ flights~\cite{coutu:muons}, 
with only minor modifications for differing instrument configuration.  
The results are shown in Figures~\ref{fig:flux_gnd}, ~\ref{fig:flux_atm}, and 
\ref{fig:growth}, and listed in Table~\ref{tbl:fluxes}.  The differential 
intensities as a function of rigidity and atmospheric depth are shown multiplied 
by $R^{2}$ for better comparison.  Also shown are the results of other 
measurements (\cite{boezio:muons, bellotti:muons, motoki:muons}).  
The level of agreement is quite good despite the differing solar epochs and 
rigidity cutoffs.  The curves shown are for calculations with the 
TARGET algorithm with the same parameters as for Figure~\ref{fig:ratiofs}. 

The agreement between the HEAT measurements and model predictions is for the 
most part excellent, except for the very highest altitudes where the predictions 
are below the measured intensities. We note that in Fig.~\ref{fig:flux_gnd} there
appears to be a tendency for the measurements of the ground $\mu^-$ intensities made
at a 4.5~GV geomagnetic cutoff to be slightly higher than those at sub-GV cutoff.
We do not expect this effect to be real, as energetic muons ($R>1$~GV) at the ground 
are the end products of parent cosmic rays well above the geomagnetic cutoff at 
either location. We note that differential intensities are rescaled by $R^2$, so that 
the effect of any small systematic error in the rigidity determination is magnified.

\begin{table*}
\caption{Atmospheric $\mu^{-}$ intensities [in (cm$^{2}$ s sr GeV/c)$^{-1}$] 
and number of recorded events as a function of rigidity and atmospheric
depth.}
\begin{center}
\footnotesize
\begin{tabular}{cccccc}
\hline\hline
 & 4--7 g/cm$^{2}$ & 7--13 g/cm$^{2}$ & 13--32 g/cm$^{2}$ & 32--67
g/cm$^{2}$ & 67-140 g/cm$^{2}$ \rule{0mm}{4mm} \\
\hline
0.3--0.5 & $(1.51 \pm 0.26)\times 10^{-3}$ & $(2.06 \pm 0.31)\times
10^{-3}$ & $(4.62 \pm 0.61)\times 10^{-3}$ & $(8.5 \pm 1.1)\times 10^{-3}$
& $(1.18 \pm 0.15)\times 10^{-2}$ \rule{0mm}{4mm} \\
GV & 61 & 108 & 279 & 415 & 672 \\
0.5-0.8 & $(1.23 \pm 0.18)\times 10^{-3}$ & $(1.58 \pm 0.21)\times
10^{-3}$ & $(3.27 \pm 0.41)\times 10^{-3}$ & $(6.14 \pm 0.74)\times
10^{-3}$ & $(8.82 \pm 1.09)\times 10^{-3}$ \\
GV & 110 & 186 & 440 & 670 & 1116 \\
0.8--1.2 & $(5.21 \pm 0.85)\times 10^{-4}$ & $(1.02 \pm 0.14)\times
10^{-3}$ & $(1.84 \pm 0.24)\times 10^{-3}$ & $(3.61 \pm 0.44)\times
10^{-3}$ & $(5.58 \pm 0.69)\times 10^{-3}$ \\
GV & 71 & 185 & 380 & 607 & 1086 \\
1.2-1.7 & $(4.08 \pm 0.67)\times 10^{-4}$ & $(5.26 \pm 0.76)\times
10^{-4}$ & $(1.12 \pm 0.15)\times 10^{-3}$ & $(2.32 \pm 0.29)\times
10^{-3}$ & $(3.10 \pm 0.39)\times 10^{-3}$ \\
GV & 71 & 120 & 294 & 498 & 768 \\
1.7-2.5 & $(1.42 \pm 0.27)\times 10^{-4}$ & $(2.75 \pm 0.41)\times
10^{-4}$ & $(5.63 \pm 0.75)\times 10^{-4}$ & $(1.18 \pm 0.15)\times
10^{-3}$ & $(1.55 \pm 0.20)\times 10^{-3}$ \\
GV & 41 & 104 & 246 & 415 & 629 \\
2.5--4.0 & $(6.7 \pm 1.3)\times 10^{-5}$ & $(1.01 \pm 0.16)\times 10^{-4}$
& $(2.10 \pm 0.29)\times 10^{-4}$ & $(4.63 \pm 0.59)\times 10^{-4}$ &
$(6.34 \pm 0.81)\times 10^{-4}$ \\
GV & 37 & 73 & 174 & 311 & 494 \\
4.0--8.0 & $(3.11 \pm 0.59)\times 10^{-5}$ & $(4.13 \pm 0.67)\times
10^{-5}$ & $(8.0 \pm 1.1)\times 10^{-5}$ & $(1.43 \pm 0.19)\times 10^{-4}$
& $(1.88 \pm 0.24)\times 10^{-4}$ \\
GV & 43 & 75 & 165 & 239 & 363 \\
8.0--16.0 & $(6.6 \pm 1.8)\times 10^{-6}$ & $(9.9 \pm 2.0)\times 10^{-6}$
& $(1.62 \pm 0.28)\times 10^{-5}$ & $(2.87 \pm 0.45)\times 10^{-5}$ &
$(4.01 \pm 0.58)\times 10^{-5}$ \\
GV & 17 & 34 & 63 & 90 & 145 \\
16.0--32.0 & $(1.52 \pm 0.64)\times 10^{-6}$ & $(1.68 \pm 0.62)\times
10^{-6}$ & $(4.3 \pm 1.0)\times 10^{-6}$ & $(6.1 \pm 1.4)\times 10^{-6}$ &
$(7.2 \pm 1.4)\times 10^{-6}$ \\
GV & 6 & 8 & 23 & 25 & 39 \\
32.0--50.0 & & $(4.1 \pm 4.1)\times 10^{-7}$ & $(3.7 \pm 3.7)\times
10^{-7}$ & $(5.4 \pm 5.4)\times 10^{-7}$ & $(3.0 \pm 1.1)\times 10^{-6}$
\\
GV & 0 & 1 & 1 & 1 & 9 \\
\end{tabular}
\begin{tabular}{cccccc}
\hline
 & 140-250 g/cm$^{2}$ & 250-350 g/cm$^{2}$ & 350-840 g/cm$^{2}$ & 840-960
g/cm$^{2}$ \rule{0mm}{4mm} \\
\hline
0.3--0.5 & $(1.88 \pm 0.24)\times 10^{-2}$ & $(1.40 \pm 0.19)\times
10^{-2}$ & $(6.69 \pm 0.88)\times 10^{-3}$ & $(2.28 \pm 0.31)\times
10^{-3}$ \rule{0mm}{4mm} \\
GV & 477 & 213 & 247 & 172 \\
0.5-0.8 & $(1.29 \pm 0.16)\times 10^{-2}$ & $(9.7 \pm 1.2)\times 10^{-3}$
& $(5.51 \pm 0.68)\times 10^{-3}$ & $(1.64 \pm 0.21)\times 10^{-3}$ \\
GV & 730 & 329 & 457 & 283 \\
0.8--1.2 & $(8.4 \pm 1.1)\times 10^{-3}$ & $(6.83 \pm 0.86)\times 10^{-3}$
& $(3.97 \pm 0.49)\times 10^{-3}$ & $(1.39 \pm 0.18)\times 10^{-3}$ \\
GV & 737 & 359 & 509 & 373 \\
1.2-1.7 & $(4.84 \pm 0.62)\times 10^{-3}$ & $(4.37 \pm 0.56)\times
10^{-3}$ & $(2.74 \pm 0.34)\times 10^{-3}$ & $(1.22 \pm 0.15)\times
10^{-3}$ \\
GV & 538 & 291 & 448 & 421 \\
1.7-2.5 & $(2.72 \pm 0.35)\times 10^{-3}$ & $(2.23 \pm 0.29)\times
10^{-3}$ & $(1.47 \pm 0.18)\times 10^{-3}$ & $(8.01 \pm 0.99)\times
10^{-4}$ \\
GV & 501 & 247 & 396 & 457 \\
2.5--4.0 & $(1.00 \pm 0.13)\times 10^{-3}$ & $(8.2 \pm 1.1)\times 10^{-4}$
& $(7.57 \pm 0.95)\times 10^{-4}$ & $(4.37 \pm 0.54)\times 10^{-4}$ \\
GV & 348 & 171 & 392 & 482 \\
4.0--8.0 & $(3.34 \pm 0.45)\times 10^{-4}$ & $(2.67 \pm 0.38)\times
10^{-4}$ & $(2.69 \pm 0.34)\times 10^{-4}$ & $(1.77 \pm 0.22)\times
10^{-4}$ \\
GV & 288 & 138 & 349 & 491 \\
8.0--16.0 & $(5.89 \pm 0.93)\times 10^{-5}$ & $(5.27 \pm 0.95)\times
10^{-5}$ & $(6.20 \pm 0.87)\times 10^{-5}$ & $(4.64 \pm 0.62)\times
10^{-5}$ \\
GV & 95 & 51 & 152 & 230 \\
16.0--32.0 & $(9.5 \pm 2.4)\times 10^{-6}$ & $(8.4 \pm 2.6)\times 10^{-6}$
& $(9.0 \pm 1.9)\times 10^{-6}$ & $(1.15 \pm 0.18)\times 10^{-5}$ \\
GV & 21 & 12 & 30 & 93 \\
32.0--50.0 & $(1.9 \pm 1.4)\times 10^{-6}$ & $(3.9 \pm 2.3)\times 10^{-6}$
& $(3.0 \pm 1.4)\times 10^{-6}$ & $(2.66 \pm 0.83)\times 10^{-6}$ \\
GV & 2 & 3 & 5 & 12 \\
\hline\hline
\end{tabular}
\normalsize
\end{center}
\label{tbl:fluxes}
\end{table*}

\section{Conclusions}

We have measured with good statistics the muon charge ratio 
$\mu^{+}/\mu^{-}$ as a function of atmospheric depth in the momentum 
interval 0.3--0.9 GeV/c and the differential $\mu^{-}$ intensities in the 
0.3--50 GeV/c range and for atmospheric depths between 4--960 g/cm$^{2}$. 
We have found that our charge ratio is $\sim$1.1 for all atmospheric
depths and is consistent, within errors, with other measurements and the
model predictions.  We have found that our measured $\mu^{-}$ intensities 
are also largely consistent with other measurements, and with the model predictions,
despite varying solar epochs and geomagnetic rigidity cutoffs. This reinforces
the conclusion of \cite{coutu:muons} that model calculations are adequate to
the task of predicting neutrino fluxes. The discrepancy between our
measurements and the model predictions at shallow atmospheric depths is of 
little import to our understanding of neutrino fluxes, as the bulk of the 
neutrinos come from decay processes much deeper in the atmosphere.

\begin{acknowledgments}
This work was supported by NASA Grants No. NAG 5-5058, No. NAG 5-5220, 
No. NAG 5223, and No. NAG 5-2230, and by financial assistance from our 
universities.  We are grateful to T. Stanev and T. K. Gaisser for sharing
with us their TARGET Monte Carlo algorithm.  We wish to thank the National 
Scientific Balloon Facility and the NSBF launch crews for their excellent 
support of balloon missions, and we acknowledge contributions from 
A. S. Beach, D. Kouba, M. Gebhard, S. Ahmed, P. Allison, and S. Verbovsky.
\end{acknowledgments}


\end{document}